\documentclass[11pt]{article}
\pdfoutput=1

\usepackage[final]{acl} % Use "final" or "preprint" as needed
\usepackage{times}
\usepackage{latexsym}
\usepackage[T1]{fontenc}
\usepackage[utf8]{inputenc}
\usepackage{microtype}
\usepackage{inconsolata}
\usepackage{graphicx}

\title{Will AI shape the way we speak? \\ The emerging sociolinguistic influence of synthetic voices}
\author{
Éva Székely\textsuperscript{1} \qquad
Jūra Miniota\textsuperscript{1} \qquad
Míša (Michaela) Hejná\textsuperscript{2} \\
\textsuperscript{1}Department of Speech, Music and Hearing, KTH Royal Institute of Technology, Sweden \\
\textsuperscript{2}Department of English, Aarhus University, Denmark \\
\texttt{szekely@kth.se, jura@kth.se, misa.hejna@cc.au.dk}
}
\begin{document}
\maketitle

% Abstract
\begin{abstract}
The growing prevalence of conversational voice interfaces, powered by developments in both speech and language technologies, raises important questions about their influence on human communication. While written communication can signal identity through lexical and stylistic choices, voice-based interactions inherently amplify socioindexical elements—such as accent, intonation, and speech style -- which more prominently convey social identity and group affiliation.
There is evidence that even passive media such as television is likely to influence the audience's linguistic patterns. Unlike passive media, conversational AI is interactive, creating a more immersive and reciprocal dynamic that holds a greater potential to impact how individuals speak in everyday interactions. Such heightened influence can be expected to arise from phenomena such as acoustic-prosodic entrainment and linguistic accommodation, which occur naturally during interaction and enable users to adapt their speech patterns in response to the system.
While this phenomenon is still emerging, its potential societal impact could provide organisations, movements, and brands with a subtle yet powerful avenue for shaping and controlling public perception and social identity. We argue that the socioindexical influence of AI-generated speech warrants attention and should become a focus of interdisciplinary research, leveraging new and existing methodologies and technologies to better understand its implications.
\end{abstract}

% Main Sections
\section{Introduction}
This position paper proposes that the increasing scale and quality of verbal interactions with AI has the potential to influence people’s habitual voice and speaking style on an unprecedented scale. Recent advancements in large language models (LLMs) and text-to-speech (TTS) technology now enable realistic, expressive, human-like conversations. Moreover, breakthroughs in conversational AI systems, such as naturalistic turn-taking \cite{arora2025talking} and interruption handling \cite{cao2025interruption}, are expected to drastically increase the scale of spoken interactions with AI.
While both written and spoken language can convey aspects of identity, they do so through different channels. In writing -- especially in informal settings -- word choice, grammar, and style can reflect social traits such as age, gender, or cultural affiliation (e.g., \citealt{Rubin1995}). However, spoken interaction inherently and therefore unavoidably conveys such extralinguistic traits through the voice itself. This means that the societal impacts of increased voice-based interactions with AI are likely to differ considerably from those of text-based interactions.

\section{Socioindexicality in spoken AI interaction}
\subsection{Spoken language and social identity}
One particularly relevant concept in this context is socioindexicality, which refers to how features of communication signal social identity and group affiliation \cite{Silverstein2003, eckert2019limits}.
In spoken language, socioindexical elements, such as accent, intonation, and speech style,  play a crucial role in conveying these social cues. A wide range of identity-related aspects -- including personality and wellbeing -- can be signalled, and indexed, through linguistic variation, including phonetic and phonological variation (e.g. \citealt{CamKib2009, Pharao2014, PodesvaCallier2015, PaladinoMazzurega2019, Guyetal2022, HopeLilley2023, Grammon2024}).

Given the increasing realism and human-likeness of synthetic voices, socioindexical elements embedded in AI-generated speech may extend the role of conversational AI beyond functionality. These elements could potentially become socially influential, producing tangible effects on users' perceptions and behaviors through specific vocal traits.

\subsection{Acoustic-prosodic entrainment and linguistic accommodation}
Entrainment (also called alignment, accommodation, or convergence) refers to the tendency of dialogue partners to become more similar in their communicative behaviors \cite{levitan11_interspeech, wynn2022classifying}.
In human-human conversations, people naturally align on various levels -- choice of words, sentence structures, speech rate, intonation, etc. -- which can foster rapport \cite{miles2009rhythm}, signify cooperation \cite{pellegrino2023speakers} and reinforce social bonds between speakers.

A substantial body of work shows that humans do adjust their speech and language when interacting with machines.
Even in early studies of human–machine dialogue, researchers observed entrainment effects that parallel those found in human-human conversation. Users adapt their speech to align with artificial interlocutors in both lexical and prosodic domains. For instance, speakers converged on the vocabulary used by spoken dialogue systems \citep{parent2010lexical} and conversational agents \citep{ostrand23_interspeech}. Prosodic convergence has also been documented in interactions with animated personas
\citep{oviatt2004toward}, social robots \citep{cohn2023vocal}, and virtual tutors \citep{tsfasman2021towards}. Participants modulated features like pitch, amplitude, and speech rate to more closely match the agent’s delivery. Speakers even adjusted their speaking rate when addressing early spoken dialogue systems \citep{bell2003prosodic}, and entrained to turn-taking rhythms in expressive humanoid robots \citep{breazeal2002regulation}. 
More recent findings show that the degree of prosodic entrainment can vary based on the agent’s politeness and perceived humanness \citep{horstmann2024communication, tsfasman2021towards}.
These findings indicate that entrainment in HCI is not limited to functional adaptation, but it also reflects socially grounded mechanisms that operate similarly with both artificial and human interlocutors.

\subsection{From alignment to identity expression}
Linguistic accommodation is commonly viewed in sociolinguistics as a key mechanism that may influence how linguistic variation evolves into dialect formation and, eventually, language change \cite{hinskens2005role}. In other words, short-term accommodation during repeated conversational exchanges can, over time, lead to long-term changes both at the individual level \cite{nguyen2015role, lee2010quantification}, as well as at the community level, where it can lead to the spread and adoption of innovative linguistic variants \cite{hinskens2005role}.
Perceived prestige -- often associated with artificial intelligence -- has been shown to amplify this effect \cite{lev2014experimental}.
Linguistic accommodation being a reciprocal process, the rise of adaptive conversational AI \cite{brandt2024towards, pollmann2023entertainment} can be expected to reinforce this phenomenon even further.

This suggests that people could begin to absorb AI-influenced speech patterns in general contexts, potentially shaping their everyday language and, with it, their expression of identity.
Evidence of a similar influence is already emerging with text-based chatbots, where users adopt words or phrases commonly generated by language models and subsequently use them in their spoken language, as observed in YouTube videos \cite{Yakura2024Empirical}.

\section{Potential societal influence}
\subsection{Lessons from media}
Over the past three decades, sociolinguistic research has explored how media influences speech patterns and linguistic performance \cite{Tagli2014} and how it contributes to language change \cite{Kristiansen2014}. Studies show that exposure to media can diffuse linguistic features, both on the lexical \cite{Trudgill2014} and on the phonological level \cite{oviatt2004toward}. 
While most of the research in this broader area has targeted either written language \citep{Crystal2006, Tagli2016} or the potential effects of modes of communication such as Instant Messaging on spoken language \citep{TagliDenis2008}, one of the most notable endeavours in the area of speech influence is presented by \citet{sayers2014mediated}, who proposes a mediated innovation model to operationalise the role of media exposure and engagement on `everyday' linguistic and speech changes.

Regarding phonetic and phonological features, few sociolinguistic studies are available. One prominent example of media influence on the acoustic-phonetic level is the phonological shifts observed in Glaswegian speech linked to psychological engagement with a popular London-based TV drama \cite{stuart2013television}. The researchers found that TH-fronting and L-vocalisation can be linked to psychological engagement with characters on the \textit{EastEnders} soap opera.

\noindent Beyond linguistic variation, \citet{Kristiansen2014} explores how the media shape language change through \textit{ideology} -- that is, socially shared beliefs about which ways of speaking are desirable, appropriate, or prestigious -- and calls for further research into the media's role in shaping such perceptions.
As conversational AI becomes a more common mode of media engagement, it participates in these ideological processes, subtly reinforcing or shifting language attitudes through ongoing, interactive exposure and perceived prestige \cite{xi2024prestige}.

\subsection{The rise of an engineered language change?}
 As shown by prior research, even before the widespread adoption of AI voices, media had already demonstrated its potential to influence how people speak and express themselves -- often in ways that extend beyond direct interpersonal interaction and diffuse across distant geographic regions.
Generative AI introduces an interactive dimension that is likely to amplify such influences.
Speakers could actively -- yet often unconsciously -- incorporate socioindexical traits exhibited by conversational agents in their habitual speaking style. Through this process, companies, political movements, and other organizations may gain a new avenue for subtle influence.
By designing AI voices with specific socioindexical characteristics -- such as accents, speech styles, or voice quality features -- these actors may encourage users to adopt speech patterns that signal affiliation with a brand, ideology, or social group.
This influence could shape social identity markers and foster subconscious associations with particular movements or subcultures.
Likely emerging examples of this phenomenon include AI companions \cite{zhang2025real}, AI-powered interactive virtual influencers \cite{yu2024artificial}, and chatbot versions of human influencers.

\subsection{Societal implications and ethical risks}
It is further pertinent to ask to what extent any linguistic profiling within AI voices might  contribute to linguistic discrimination, which is an established phenomenon: linguistic variants can be and have been utilised to classify speakers into social categories and to mistreat these speakers as a result \citep{Pur1999, GlusDov2010, LippiGreen2012, KraPap2020}. 
On a more general level, AI may reinforce normative biases by defaulting to voices representative of the unmarked and commonly employed white, cis-gender, heterosexual, and able-bodied speakers. This may reinforce already existing dominant norms.
In this light, the potential influence of conversational AI on speech production is not merely a linguistic curiosity but could pose actual ethical harm \cite{hutiri2024notmyvoice}.

We believe that the potential societal impacts of AI-driven socioindexical influence on speech patterns and identities can be rather substantial.
While empirical evidence is still emerging, we identify socioindexical influence as an under-explored area with significant potential for societal impact.
Understanding this phenomenon now, while it is nascent, can be an opportunity to shape ethical design and governance before its effects become widespread. However, whether this is the case, and to what extent, remains unexplored. Considering the rapid advances in relevant technology and the widespread engagement with conversational AI, it is important to develop methods for understanding which speech characteristics may become influential or habitualised through interaction with synthetic voices.

\section{Research opportunities and challenges}
\subsection{Studying short- and long-term effects}
While short-term accommodation to AI voices in interaction is established, it remains unclear whether and how these immediate, conversation-specific adaptations carry over into one’s long-term speech habits outside the interaction. Most studies to date examine alignment within an interaction; they do not test if a person’s baseline speaking style changes after repeated exposures. 
Studying the nature of the long-term potential and topical influences requires methodologies that extend beyond but include traditional sociolinguistic approaches, particularly when considering the interactive nature of modern media and conversational systems \cite{sayers2014mediated}. 
Indeed, the still highly unique study by \citet{stuart2013television} presents a tour de force which, among other things, demonstrates the methodological complexities and challenges of investigating the role of the media on phonetic speech variation outside of a laboratory setting.
Individual variation in susceptibility to such influence should also be considered. Not all speakers will accommodate to synthetic voices the same way -- some may even actively resist alignment. Future work should explore who adapts, who resists, and why.

\subsection{Experimental approaches with TTS and Conversational AI}
The same technologies that raise questions about socioindexical influence -- speech synthesis and conversational AI -- also bring new methodological possibilities.
Advances in speech synthesis provide researchers with unprecedented control over acoustic-prosodic features, enabling experimental designs that isolate individual variables such as pitch, speech rate, and voice quality.
These systems can also affect features such as formality, allowing the development of methodologies that use TTS trained on spontaneous speech data as a research tool \cite{szekely2024voice, omahony24_interspeech}.
Moreover, recent developments in large-scale neural TTS systems trained on thousands of hours of speech have dramatically lowered the threshold for high-fidelity voice replication \cite{casanova24_interspeech}.
Fine-tuning these models on as little as ten minutes of in-the-wild speech material makes it possible to reproduce sociolects without requiring extensive recordings. In addition, zero-shot TTS and voice conversion \cite{lameris24_interspeech} enable the transfer of these speech patterns to different voice identities, which facilitates experimental comparisons across demographic categories like gender and age and even vocal characteristics. While such manipulations must be approached with care to preserve indexical plausibility and perceptual coherence \cite{seaborn2025unboxing},
this capacity for decoupling linguistic features from speaker identity expands the range of testable hypotheses in experimental sociolinguistics.

Such synthetic stimuli can be used in both perception and production studies, including shadowing tasks \citep{Laycock2021The, pardo2018comparison}, to estimate phonetic convergence to emerging sociolects. Interactive experimental designs also become feasible through research-grade conversational agents equipped with controllable TTS \cite{wang24t_interspeech}, enabling A/B testing of entrainment during dialogue.
These methods can be further complemented by sociolinguistic interviews or ethnographic observations on social media materials \cite{Yakura2024Empirical}.

\subsection{Multidisciplinary opportunities}
The complexity of media influence, which is shaped by engagement, identity, and context, calls for a transdisciplinary approach \cite{Andro2014}.
Studying the sociolinguistic impact of synthetic voices, and especially the conditions under which local adaptation might evolve into long-term language change, will require collaboration across multiple disciplines. Building research-purpose TTS and Conversational AI systems, analysing subtle language and speech variations, and interpreting social impact are key components of this research direction and will necessitate close collaboration between engineers, linguists, social scientists, and ethicists.
We anticipate that the increasing presence of speech AI in society will lead to further research areas becoming increasingly multidisciplinary.
This may require rethinking research infrastructures or even education programs.

\section{Conclusion}
This position paper calls attention to the need for a concerted effort to address the socioindexical influence of AI-generated voices in interaction.
First, it is imperative to establish the existence and extent of this emergent phenomenon.
This requires a foundational understanding of how AI interaction may impact speech patterns and identity expression among users.
Second, the development of robust methodologies is critical for systematically studying and measuring these influences.
Finally, we must begin to consider the broader implications, including ethical, societal, and legal dimensions.
Addressing these priorities will help us prepare to understand and manage the implications of voice-based conversational AI for human speech, communication, social identity, and its potential role in driving language change.

\section*{Acknowledgements}
This research is supported by the Swedish Research Council project Perception of speaker stance (VR-2020-02396), and the Riksbankens Jubileumsfond project CAPTivating (P20-0298).
We thank the anonymous reviewers for their insightful comments which helped improve the final version of this paper.
We are also deeply grateful to the many colleagues and friends who generously shared feedback on earlier drafts, including Alyssa Allen, Matthew Aylett, Jens Edlund, Emer Gilmartin, Maxwell Hope, David House, Tanya Karoli Christensen, Johannah O'Mahony, Amir H. Payberah, André Pereira, Graham Pullin, Fin Tams-Grey, Ilaria Torre and Marcin Włodarczak. Their thoughtful suggestions and encouragement were invaluable in shaping this work.

% References
%\bibliography{.bbl}

% Appendices (if any)
%\appendix
%mi\section{Appendix}
%Additional details can go here.

\end{document}